# SAXS measurements of azobenzene lipid vesicles reveal buffer-dependent photoswitching and quantitative *Z→E* isomerisation by X-rays

Martina F. Ober[1], Adrian Müller-Deku[2], Anna Baptist[1], Heinz Amenitsch[3], Oliver Thorn-Seshold[2], and Bert Nickel[1]

[1] Faculty of Physics and CeNS, Ludwig-Maximilians-Universiät München, Geschwister-Scholl-Platz 1, 80539 Munich, Germany
[2] Department of Pharmacy, Ludwig-Maximilians-Universität München, Butenandtstraße 5-13, 81377 Munich, Germany
[3] Institute of Inorganic Chemistry, Graz University of Technology, Stremayrgasse 9, 8010 Graz, Austria

Corresponding author: nickel@lmu.de

**Abstract** Photoresponsive materials feature properties that can be adjusted near-instantaneously, reversibly, and with high spatiotemporal precision, by light. There is considerable interest in maximising the degree of switching, and in measuring this degree during illumination in complex environments. We study the switching of photoresponsive lipid membranes that allow precise and reversible manipulation of membrane shape, permeability, and fluidity. Though these macroscopic photoresponses are clear, it is unclear how large the changes of *trans/cis* ratio are, and whether they can be improved. Here, we used small-angle X-ray scattering to measure the thickness of photoswitchable lipid membranes, and correlate thickness to *trans/cis* ratios. This revealed an unexpected dependency of photoswitching upon aqueous phase composition, highlighting smaller-than-expected photoswitching with deionized water, and showed thickness variations twice as large as previously observed. Soft X-rays can quantitatively isomerise photolipid membranes to the all-trans state: both enabling more powerful X-ray-based membrane control and underlining the role of high energy X-rays for observation-only soft matter experiments.

**Introduction**

Azobenzene photoswitches are molecular units, which can be switched between *trans* and *cis* isomer states at the single-molecule level by light. By embedding photoswitches into polymers or supramolecular assemblies containing many individual photoswitch units, materials with mechanical, electrical or optical properties that depend on the *trans/cis* ratio of the photoswitch population can be created: i.e., photoresponsive materials where these properties are adjustable by light.[1-6] Supramolecular assemblies can also provide further degrees of complexity, for example the populations of the two isomers may phase separate,[7] or a highly ordered assembly process may select for or stabilise one of the isomer states.[8]

As a photoresponsive material's properties depend on the isomer ratio, there is considerable interest in measuring and maximising the degree of switching between mostly-*trans* and mostly-*cis* populations.[9] Typically, it is not possible to perform quantitative photoisomerisations to all-*cis* or to all-*trans* azobenzene populations, due to absorption overlaps.[8, 10-13] Hence, illuminations lead to mixed photostationary state (PSS) populations, with *trans/cis* ratios that depend on the absorption coefficients of the two isomers and the quantum yields of their photoisomerisation. However, a rare example of quantitative switching in one direction (to all-*trans*) was pioneered by Hecht *et al.*, using electrocatalytic pathways that transiently oxidise or reduce the azobenzenes.[14-16]

Here we study the switching of photoresponsive lipid membranes assembled from azobenzene-containing phosphatidylcholine (azo-PC), a synthetic lipid with a light-responsive azobenzene group in one of its two hydrophobic tails. Optical control of membranes constructed from pure azo-PC allows precise and reversible manipulation of many mechanical properties including membrane shape, permeability, fluidity, and domain formation, and influence membrane protein function.[4, 17-20] So far, it remains unclear how large the changes of *trans/cis* ratio are that cause these property changes; and whether these ratio are close to the theoretical maxima of pure states, or whether they could be substantially improved by tailored conditions and switching stimuli - with associated improvements to switching of biophysical properties. For example, ultrastructure studies based on X-ray experiments have shown photostimulated membrane thickness changes of ca. 4 Å for ca. 42 Å azo-PC membranes:[4] but it was not known if this 10% change is already maximal, nor was it known what population-level of *trans/cis* ratios were responsible for this change. This partly derives from a technical challenge: while it is straightforward to measure *trans/cis* ratios in molecular solutions by a variety of methods, it is not straightforward to measure them in the pure azo-PC membranes, which are ordered anisotropic assemblies.

In this work, we develop small-angle X-ray scattering (SAXS) as a powerful method to measure the thickness of photoswitchable lipid membranes, and we use orthogonal measurements to relate SAXS-measured thickness to population-level *trans/cis* ratios. This revealed an unexpected sensitivity of the photoswitches inside the lipid membranes to the aqueous buffer conditions outside them, particularly with a smaller-than-expected photoswitching yield under low ionic strength. Furthermore, by mixing pre-switched monomers, we show that membrane variations of about 8 Å are experimentally possible when photoswitching efficiency is unimpeded. This is twice as large as the thickness variation observed in previous vesicle photoswitching experiments,[4] and is very much larger than the membrane thickness variation achievable by temperature variations with conventional lipids.[21-24]

We also discovered unexpectedly that while hard X-rays do not switch the membranes under study, soft X-rays (8 keV) efficiently and *quantitatively* isomerise photolipid membranes to the all-*trans* state within seconds, which we attribute to radical redox reactions following X-ray dose deposition in the medium. This enables soft X-rays to enforce a higher degree of membrane property control than photoswitching alone can achieve, while emphasising the role of high energy X-rays as low dose probes in soft matter experiments.

**Materials and Methods**

**Synthesis of azo-PC and reference photoswitch FAzoM**

Azo-PC was purchased from Avanti Polar Lipids, Inc. (Alabama, United States). Novel reference photoswitch FAzoM was synthesised by standard reactions (see Supporting Information).

**Synthesis of azo-PC and reference photoswitches for benchmarking**

In our hands, standard techniques to measure *trans:cis* ratios in molecular solutions (H-NMR, HPLC) were not reproducible when applied *in situ* to lipid membranes (ordered assemblies). Destructive readouts (e.g. adding cosolvents to homogenise membrane/water mixtures before HPLC measurement) were also tested but were also not reproducible, which we attributed to difficulties arising from the surfactant nature of the AzoPC. Finally, we developed a method to relate photolipid membrane *trans/cis* isomer ratio to the thickness determined by SAXS by using a calibration series for lipid membranes of known *trans/cis* composition. This series was created by mixing preconditioned all-*trans* and mostly-*cis* stocks (see below). The *trans/cis* ratios in the preconditioned AzoPC stocks were in turn determined by comparison of their UV-Vis spectra with that of the isoelectronic reference photoswitch FAzoM in the same conditions, since the *trans/cis* ratio of the apolar FAzoM can be reliably quantified by HPLC (see Supporting Information section S1 for details).

**Preconditioning and mixing of azo-PC**

Azo-PC was dissolved in chloroform (25 mg/mL) and stored at -20 °C until further use. Azo-PC stocks have been stored in dark for several days to reach the all-*trans* state. After illumination of molecularly-dissolved azo-PC by UV-A LED (*Roschwege Star-UV365-03-00-00*, λ = 365 nm, 9 nm FWHM, Conrad Electronic SE, Germany) the photostationary state ($PSS_{UV}$) has a *cis* fraction of 83% (see Supporting Notes for further detail). To adjust the *cis* fraction in the assembly, we mix azo-PC in the all-*trans* state with azo-PCs with *cis* fraction of 83%, in appropriate proportions. After mixing, we follow the protocol for vesicle preparation. All preparation steps were made in the dark.

**Small unilamellar vesicle (SUV) preparation**

The Azo-PC chloroform stock solution was evaporated under a nitrogen stream and stored under vacuum for 12 h. The resulting dry lipid film was dissolved in cyclohexane, and exposed to a vacuum of 6 x $10^{-3}$ mbar at a temperature of – 60 °C yielding a fluffy lipid powder. Immediately after lyophilisation, the azo-PC powder was hydrated with deionized (DI) water (Milli-Q, Reptile Bioscience Ltd., Boston, MA), or with PBS buffer (pH 7.5), or with 1x TE buffer (10 mM Tris, 1 mM EDTA, pH 8), to a final concentration of 30 mg/mL. The suspension was gently vortexed and subjected to five freeze/thaw cycles. Finally, the sample solution was extruded ca. 25 times through a polycarbonate membrane with a pore diameter of 50 to 80 nm using a Mini Extruder (Avanti Polar Lipids, Inc., Alabama, United States).

**UV-A/blue light illumination**

For photoswitching of azo-PC membranes during SAXS, we built a dual UV-A and blue light LED setup shown in the Supporting Information section S2. For UV-A illumination, we focus a high-power LED (*Roschwege Star-UV365-03-00-00*, λ = 365 nm, 9 nm FWHM, Conrad Electronic SE, Germany) on the sample capillary. The total maximum optical power of 170 mW and a focal spot size of 4 mm$^2$ yield an irradiance of 4.25 W cm$^{-2}$. For blue light illumination, fed in by a dichroic mirror, a high-power LED (*Roschwege LSC-B*, λ = 465 nm, 18 nm FWHM, Conrad Electronic SE, Germany) was used. The blue light is focused with the same focal spot size and a total maximum optical power of 120 mW, resulting in an irradiance of 3.0 W cm$^{-2}$. The LEDs and the X-ray detector were remote controlled by TTL signals from an Arduino microprocessor (Reichelt electronics GmBH & Co. KG, Germany).

**SAXS measurements at 17.4 keV**

X-ray data from azo-PC SUVs with preconditioned *trans:cis* ratios were recorded at a Mo-sourced small angle X-ray scattering (SAXS) setup.[25] The Mo anode delivers a beam with an energy of 17.4 keV, a beam size of 1.0 mm², and a flux of $2 \cdot 10^6$ cts·s$^{-1}$mm$^{-2}$. X-ray data were recorded by a Dectris Pilatus 3R Detector with 487 x 619 pixels of size 172 x 172 µm². All in-house SAXS measurements were performed in darkness.

**SAXS measurements at 8 keV**

SAXS data from azo-PC SUVs were recorded at the Austrian SAXS beamline at ELETTRA synchrotron using a beam energy of 8 keV[26] and a beam size of 0.5 x 2.0 mm$^2$. The sample solution was loaded in 1.5-2 mm diameter quartz glass capillaries by flow-through and placed in our UV-A/blue LED setup. A Pilatus detector from Dectris Ltd., Switzerland with 981 × 1043 pixels of size 172 × 172 µm$^2$ served as detector.

**SAXS measurements at 54 keV**

The high energy SAXS data from azo-PC SUVS were recorded at beamline P21.1 at the PETRA III ring at DESY. We measured with a beam energy and size of 54 keV and 1 x 1 mm², respectively. For the high energy SAXS experiments, the sample solution was first switched optically and then loaded in Kapton tubes of 40 mm in length and 2.5 mm in diameter. A Lambda detector (X-Spectrum GmbH, Germany) with 772 x 516 pixels with 55 x 55 µm² pixel size was used.

**Results and discussion**

Azo-PC populations are driven towards *cis*-rich photostationary state isomer mixes by UV-A light at 365 nm wavelength, and towards *trans*-rich isomer mixes by blue light of 465 nm (Fig. 1a). The absorption spectra of these mixed states, either in molecular solution or in assemblies, have been reported by us and others[4, 17-18] and substantial differences between them have been noted.[18] Previously, we have used SAXS to determine the membrane thickness for blue or UV light photostationary states reached in lipid vesicles in DI water [4]. Here, we prepared calibration series of azo-PC vesicles of known *trans/cis* ratios by mixing stocks of dark-adapted all-*trans* azo-PC with 83% *cis* azo-PC (see methods and SI), using calibrants at 0%, 10%, 19%, 39%, 58% and 83% *cis* isomers to cover the full range of ratios accessible during membrane photoswitching, with an estimated error margin of ±5% of the given value to account for mixing precision as well as *cis* fraction uncertainties in the preconditioned samples. The referencing method developed to determine the stock *cis* percentage is, as far as we know, novel in this field; and it can prove generally useful for analysis of systems where measurements in bulk (e.g. membrane UV-Vis spectra) cannot be interpreted with reference to more easily characterised molecular measurements (e.g. UV-Vis spectra in monomer solution), see Methods and Supporting Information.

The SAXS measurements for these calibrant ratios are summarized in Fig. 1b. The X-ray intensity distributions are typical for lipid bilayer samples.[27-28] The distributions vary strongly with the *trans*-to-*cis* ratio, e.g. the intensity dip around $q = 0.05\ \text{Å}^{-1}$ shifts consecutively to higher q-values with increasing *cis* content. This indicates that the membrane thickness decreases with increasing *cis* isomer fraction. To extract the head-to-head distances $d_{HH}$, we model the SAXS intensities in Fig. 1b by an established lipid bilayer electron density profile, using the software SasView[29] (see supporting information section S5). The results of this analysis are shown in Fig. 1c. The dark-adapted state shows a head-to-head-distance of $d_{HH}(0 \pm 5\%\ cis) = 40.9\ \pm$ 0.6 Å. The analysis reveals that the thickness of the azo-PC membrane depends almost linearly on the *trans*-to-*cis* ratio over a broad range. Increasing the percentage of *cis* azo-PC thins the membrane down to $d_{HH}(83 \pm 5\% \, cis) = 33.0 \pm 0.5 \, \text{Å}$ for 83% *cis*. Thus, the membrane thickness changes by 8 Å.

In the following measurements, we us this relation between membrane thickness and *cis* fraction to infer the *cis* fraction of various photostationary states in response to illumination and buffer conditions (Fig. 1). Our previous photoswitching experiments have been performed with vesicles in deionized water (DI water): a common choice for lipids since DI water facilitates unilamellar membrane formation by increasing lipid vesicle stability. The maximal membrane thickness change obtained by optical switching under these conditions was only 4 to 5 Å. The new experiments with premixed lipids reported in Fig. 1 now reveal that this optical control window covers only about half of the thickness change effect which could be achieved in ideal photoswitching conditions. We therefore studied two widely used buffer systems, phosphate-buffered saline (PBS), and a mixture of Tris with EDTA (1xTE), for their influence on the optical control window. The analysis of the SAXS data is condensed in Fig. 2, the full data set is shown in Fig. S3 and Fig. S4a.

The dark-adapted states (all-*trans*) yielded mean membrane thicknesses of $d_{HH}$ (dark, buffers) = 42.9±0.2 Å and $d_{HH}$ (dark, deionized water) = 41.9±0.9 Å (Fig 2a+b, circles, solid line). Hence all conditions allow for the formation of dense, all-trans membranes in the dark. To photoswitch towards high *cis* contents, UV-A illumination was used. Independent of the usage of buffer or not, the UV photostationary states do not reach a *cis* fraction of 83% as obtained by preconditioning monomers in chloroform. Instead, the observed membrane thicknesses of $d_{HH}^{UV} = 34.8 \pm 0.6 \, \text{Å}$ (see Fig 2b, dashed line and Fig. S4) indicates *cis* contents around 64%. This is not unexpected, given the blue-shifting of azobenzene absorption spectra seen upon going from molecular solutions to assembled systems.[18]

However, when using blue light to photoswitch from a *cis*- to a *trans*-enriched state, the outcomes depend greatly on whether the system is buffered or not. In DI water, the maximal blue light photostationary state membrane thickness is $d_{HH}^{blue} = 39.0 \pm 0.3 \, \text{Å}$ (Fig 2b, squares, dotted line), i.e. 30% *cis* isomer still remains. Instead, in PBS or 1x TE buffer, the membrane thickness increases to $d_{HH}^{blue} = 42.3 \pm 0.4 \, \text{Å}$ (Fig 2a, squares, dotted line), i.e. only 3% *cis* isomer remains. Optical control of vesicles is therefore highly efficient in these buffered aqueous solutions. Experiments with simple NaCl solutions are ongoing, with results so far indicating that ionic strength is key to efficient photoswitching towards *trans*-enriched states in lipid vesicles: an effect that is not precedented for these molecules in molecular solutions.

Next, we explored methods to improve the limited optical control in DI water. For this purpose, we used X-ray exposures to deposit a certain X-ray dose into the vesicle solution, i.e. a certain energy of absorbed X-rays per mass of the exposed sample.[30] The X-ray dosing experiment was conducted as follows. First, we prepared the vesicles in a high *cis* fraction by UV-A illumination. We exposed the sample to a rather high X-ray energy of 54 keV for 60 s. Five such consecutive exposures yield an identical SAXS signal, i.e. there is no sign of any hard X-ray induced effects after 5 min in total, even though this experiment was performed at a high brilliance Petra III synchrotron beamline with full beam on the sample, cf. Fig 3a. This finding is in agreement with the fact that the cross section for photoabsorption is dramatically reduced for high X-ray energies.

Depositing high doses in water requires soft X-rays (discussion in the SI section S6). Such soft X-rays give rise to radiolysis of water, and oxidising and reducing radicals and reactive species,[31-32] which may provide pathways for catalytic redox-based unidirectional switching towards the thermodynamic groundstate (all-*trans*).[15] Therefore, we tested also the change of membrane thickness during six consecutive 8 keV X-ray exposures each of 2 s (Fig. 3a). After each X-ray exposure, a shift of the SAXS pattern was visible. To quantify the thickness change of the photomembrane, we modelled to extract the head-to-head distances, which increased dramatically after each soft X-ray exposure, depending on the total delivered X-ray dose. Remarkably, the final mean membrane thickness of $d_{HH}^{xray} = 42.5 \pm 0.3$ Å (Fig. 3b, star and square datapoints at high dose, dashed line; Fig. S4b) matches the thickness of the dark-adapted all-*trans* state $d_{HH}^{dark} = 41.9 \pm 0.9$ Å (Fig. 2b, circular data points from repeated measurements, solid line). This is a direct indication of quantitative switching obtained by soft X-ray exposure.

These soft X-ray induced changes are fully reversible (isomer state of the azobenzene photoswitch) and should not be mistaken as irreversible radiation damage. They can however be employed as a reversible X-ray dose readout. To demonstrate this, we plot the membrane thickness as observed by hard X-ray SAXS in response to soft X-ray exposure with dose expressed in kGray [kGy] in Fig. 3b. The calculation of the dose, a routine calculation in radiation protection, is explained in the SI. Here, quantitative switching starts above 100 kGy, and saturates for an X-ray dose above 700 kGy. We propose that the measurement of azo-PC *cis* to *trans* transition for SUVs in DI water can therefore be used to read out the effective X-ray dose in a regime of up to 700 kGy. This range may help to calibrate critical doses for biological SAXS experiments which can range from 51 kGy,[33] to 400 kGy,[34] to 284 – 7056 kGy.[31]

**Conclusions and summary**

SAXS with hard X-rays provides a direct read out of the membrane bilayer thickness, which we show has a linear correlation to photolipid isomerization fraction.

The 8 Å thickness change accessible to photolipid membranes under photoswitching in buffered systems (20% of membrane thickness) is massive, and far beyond the effects achievable by temperature or temperature jumps in conventional lipids.[21-22, 24, 35] The huge membrane thickness change observed here may benefit from ordering effects in the *trans* state related to H-aggregate formation as observed in spectroscopic studies.[18] Hydrophobic matching of membrane thickness with trans-membrane proteins is a wide research field that could benefit enormously from the large thickness change effects observed here.[36] Due to the rapid development of high power UV-A LEDs, and the availability of pulsed UV-A lasers at synchrotron sources,[37] we expect that future experiments with pulsed illumination may allow membrane transition times of few ms and below. At these short timescales, the hydrodynamic coupling of the intercalated water to the photolipids may give rise to interesting transient phenomena.[21] We also show that it is possible to explore a still broader range of membrane bilayer thicknesses by mixing preconditioned photolipids before assembling membranes, rather than relying only on *in situ* photoswitching, with a correspondingly greater expected range of properties.

In buffered solutions, the blue light photostationary states (3% residual *cis*) closely approach the dark-adapted state (0% residual *cis*); i.e. photoswitching experiments in physiologically stabilized environments are efficient. However, the blue light photostationary states of vesicles in DI water (30% residual *cis*) do not approach the dark-adapted state. For this purpose, quantitative *cis*-to-*trans* switching schemes are needed. Here, we show that soft X-rays of 8 keV can induce quantitative switching for these studies, which has not been realised before. For higher X-ray energies such as 17.4 keV and 54 keV, we do not observe such catalytic backswitching, which also highlight the versatility of high energy X-rays as low-dose probes. The energy dependent X-ray dose for samples with high water fraction has a minimum at 36 keV (Fig. S6), whereas the achievable SAXS signal is almost independent of the chosen beam energy (Fig. S6, inset). This suggests a favourable energy regime of 30 to 42 keV for SAXS experiments on radiation damage sensitive samples (further discussion in section S6 of the supporting information). Our experiments demonstrate that high quality SAXS data may be obtained even for weakly scattering biological samples.

**Author Contributions**

MFO performed vesicle preparation, and SAXS experiments and X-ray data analysis, and wrote the manuscript; AMD performed synthesis of azo-PC and FAzoM, and spectroscopic and HPLC analyses of *cis/trans* isomer ratios; AB developed a protocol for photolipid SUV preparation; HA enabled the SAXS experiments at ELETTRA and provided in depth expertise on SAXS calibration and data conversion; OTS designed *cis/trans* ratio measurements and wrote the manuscript; BN designed the study, joined the X-ray experiments, discussed data interpretation, and wrote the manuscript.

**Acknowledgements**

We acknowledge financial support by the German Research Foundation (DFG) through SFB1032 (Nanoagents) projects A07 (to B.N.) and B09 (to O.T.-S.) number 201269156, SFB TRR 152 project P24 number 239283807 to O.T.-S., Emmy Noether grant TH2231/1-1 to O.T.-S., and SPP 1926 project number 426018126 to O.T.-S.; by the BMBF grant no. 05K19WMA (LUCENT) to B.N.; and by the Bavarian State Ministry of Science, Research, and Arts grant "SolarTechnologies go Hybrid (SolTech)" to B.N.. This work benefited from SasView software, originally developed by the DANSE project under NSF award DMR-0520547. Portions of this research were carried out at the light source PETRA III at DESY, a member of the Helmholtz Association (HGF). We would like to thank Oleh Ivashko for assistance in using beamline P21.1 for high energy experiments.

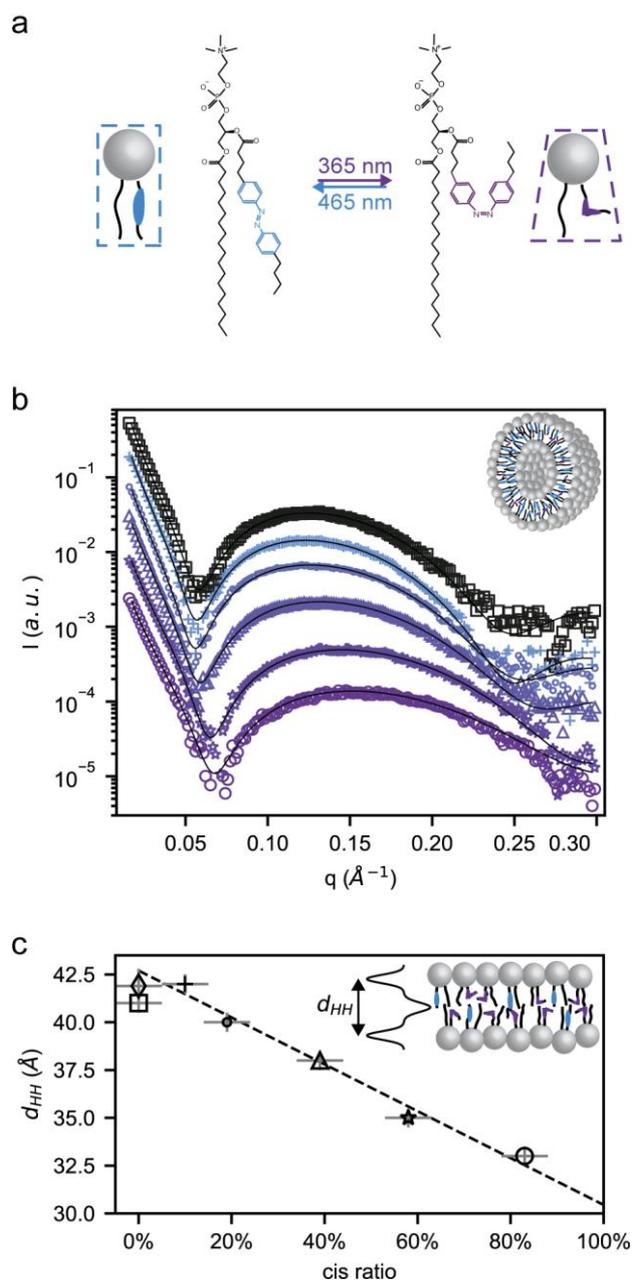

Figure 1 **X-ray structural data of photoswitchable membranes.** (a). Schematic representation of the chemical structure of the azo-PC photolipids used in this study and their *trans* and *cis* configuration. (b) SAXS intensities for unilamellar azo-PC vesicles prepared from predefined *trans* to *cis* ratios: Dark-adapted state (100:0), (90:10), (81:19), (61:39), (42:58), and (17:83) are shown as squares, crosses, dots, triangles, stars, and circles. Intensities are vertically offset for clarity. (c.) Head-to-head distance ($d_{HH}$) of azo-PC membranes as function of the percentage of azo-PCs in *cis* isomerization state are shown with the same symbols as in (b). Additionally, the mean value of $d_{HH}$ for the dark–adapted photostationary states obtained from azo-PC SUVs in DI water (Fig. 2b) is shown as diamond. The linear fit of $d_{HH}$ in dependence of *cis* isomer percentage is indicated as dashed line. See supporting information (section S7) for details.

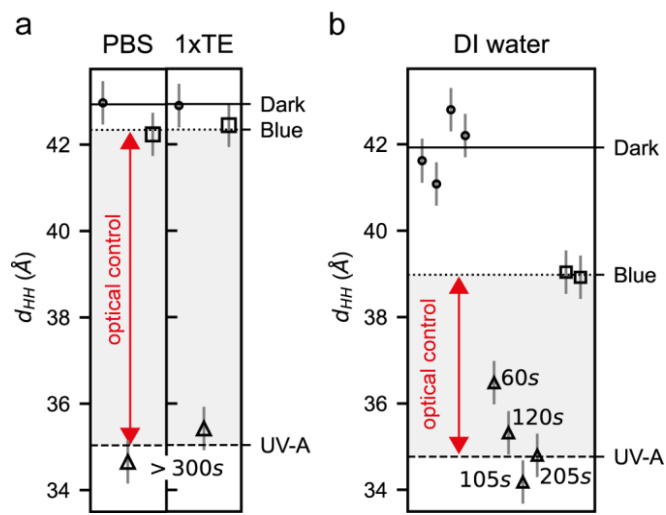

Figure 2 **Switching behaviour of photomembranes in buffer and in DI water**. (a) Head-to-head distances ($d_{HH}$) obtained for azo-PC SUVs in PBS and 1xTE buffer. (b) $d_{HH}$ obtained for azo-PC SUVs in DI water. The dark-adapted state, and several photostationary states induced via UV-A and blue light are labelled accordingly. Horizontal lines indicate mean values of $d_{HH}$ for the dark–adapted, UV-A, and blue light photostationary states, shown as solid, dashed, and dotted line, respectively. The exposure time of UV-A light while approaching the photostationary state is indicated in seconds. The optical control window for the different buffer conditions is indicated by a red double-headed arrow.

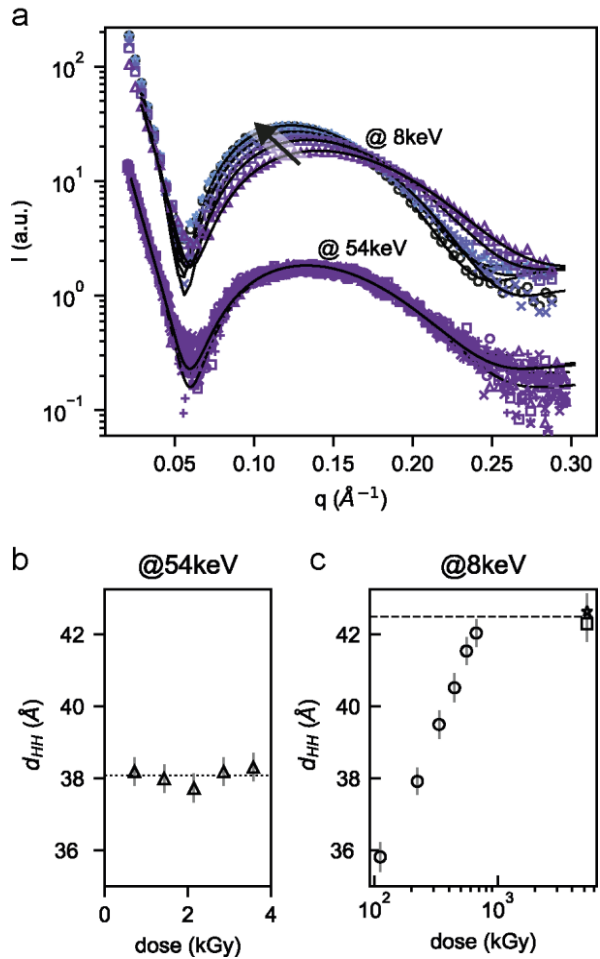

Figure 3 **Catalytic switching of photomembranes induced by soft X-rays**. (a) After initial UV illumination, two different X-ray SAXS experiments are performed at 8keV and 54keV on unilamellar photolipid vesicles. Solid curves indicate modelling of the SAXS data by symmetrical bilayers. (b) Head-to-head distances ($d_{HH}$) obtained for azo-PC SUVs in DI water as function of X-ray dose measured at 54 keV. Dotted line indicates the mean value of $d_{HH}$ for the stable UV-A induced PSS. (c) $d_{HH}$ obtained for azo-PC SUVs in DI water as function of X-ray dose measured at 8 keV. Horizontal dashed line indicate the average $d_{HH}$ after saturating 8 keV X-ray exposure.